\documentclass[12pt]{article}
\usepackage{amssymb,latexsym, amsmath,color}
\usepackage{graphicx}
\usepackage{epsfig, picinpar}
\newcommand{\rem}[1]{}
\newcommand{\url}[1]{}

\def\tfrac#1{\frac{ #1}}
\begin{document}

\title{Peakons }
\author{ Darryl D. Holm
\\
\small Mathematics Department
\\
\small Imperial College London
\\ 
\small SW7 2AZ, UK
\\ {\footnotesize email: d.holm@imperial.ac.uk} 
}
\date{}
\maketitle

\noindent
DD Holm, Peakons.
{\it Encyclopedia of Mathematical Physics}, {\bf4} (2006) 12-20.
\\ Eds. J.-P. Francoise, G.L. Naber and S.T. Tsou,  Oxford: Elsevier. 
\\ (ISBN 978-0-1251-2666-3) 

\rem{
\begin{abstract}
\noindent
PACS numbers:   47.10.+g,  52.65.Kj, 45.20.Jj, 47.65.+a, 02.40.Yy, 11.10.Ef,
05.45.-a
\end{abstract} 
}


\rem{
\noindent
{\bf Outline:}
\\ $\surd$
Shallow water background
\\ $\surd$
Singular solutions -- peakons, $b=2$ \cite{CaHo1993}
\\  $\surd$
Singular solutions -- pulsons, $b=2$ any $g$ \cite{FrHo2001}
\\  $\surd$
Singular solutions -- pulsons, any $b>1$ any $g$ \cite{HoSt2003a}
\\ $\surd$ 
Generalize the dispersionless CH equation to higher dimensions and
explain  the canonical Hamiltonian reduction of its singular momentum
solutions as a momentum map.
\\ $\surd$
Euler-Poincar\'e (EP) variational principle and EP equation for $b=2$ 
\cite{HoMaRa1998a}
\\ $\surd$ Geodesic flows (borrowed from the geometry of soliton equations and
ideal fluid dynamics) \cite{Ar1966}
\\ $\surd$ Singular solutions and momentum maps for the family of EPDiff
equations describing shallow water waves and geodesic flows in higher
dimensions
\cite{HoSt2003,HoSt2004,HoMa2004}
\\ $\surd$ EPDiff is the Euler-Poincar\'e equation for geodesic flow on
TDiff$(M)$. 
\\ --
Lie-Poisson Hamiltonian form of EPDiff
}

\paragraph{Keywords:} Solitons, Peakons, Hamilton's principle, Integrable
Hamiltonian systems, Inverse Spectral Transform (IST), Singular solutions,
Shallow water waves, Euler's equations, Geodesic motion, Diffeomorphisms,
Sobolev norms, Momentum maps, Compactons


\begin{abstract}
The peakons discussed here are singular solutions of the dispersionless Camassa-Holm (CH)
shallow water wave equation in one spatial dimension. These are reviewed in
the context of asymptotic expansions and Euler-Poincar\'e variational
principles. The dispersionless CH equation generalizes to the EPDiff
equation, whose singular solutions are peakon wave fronts in higher
dimensions. The reduction of these singular solutions of CH and EPDiff
to canonical Hamiltonian dynamics on lower dimensional sets may be understood,
by realizing that their solution ansatz is a momentum map, and momentum maps
are Poisson.
\end{abstract}

\section{Introduction}

Camassa and Holm \cite{CaHo1993} discovered the ``peakon'' solitary
traveling wave solution for a shallow water wave, 
\begin{equation}\label{singlepeakon-soln}
u(x,t)=ce^{-|x-ct|/\alpha}
\,,
\end{equation}
whose fluid velocity $u$ is a function of position $x$ on the real line and
time $t$.  The peakon traveling wave moves at a speed equal to its maximum
height, at which it has a sharp peak (jump in derivative).  Peakons are an
emergent phenomenon, solving the initial value problem for a partial
differential equation derived by an asymptotic expansion of Euler's equations
using the small parameters of shallow water dynamics. Peakons are {\it
nonanalytic} solitons, which superpose as 
\begin{equation}\label{peakontrain-soln}
u(x,t)=\sum_{a=1}^Np_a(t)e^{-|x-q_a(t)|/\alpha}
\,,
\end{equation}
for sets $\{p\}$ and $\{q\}$ satisfying canonical Hamiltonian dynamics. 
Peakons arise for shallow water waves in the limit of zero linear
dispersion in one dimension.  Peakons satisfy a partial differential equation
(PDE) arising from Hamilton's principle for geodesic motion on the smooth
invertible maps (diffeomorphisms) with respect to the $H^1$ Sobolev norm of
the fluid velocity. Peakons generalize to higher dimensions, as well. We
explain how peakons were derived in the context of shallow water asymptotics
and describe some of their remarkable mathematical properties. 

\section{Shallow water background for peakons}

Euler's equations for irrotational incompressible ideal fluid motion under
gravity with a free surface have an asymptotic expansion for shallow water
waves that contains two small parameters, $\epsilon$ and $\delta^2$, with
ordering $\epsilon\ge\delta^2$. These small parameters are
$\epsilon=a/h_0$ (the ratio of wave amplitude to mean depth) and
$\delta^2=(h_0/l_x)^2$ (the squared ratio of mean depth to horizontal length,
or wavelength). 
Euler's equations are made non-dimensional by introducing $x=l_x
x'$ for horizontal position, $z=h_0 z'$  for vertical position, $t=(l_x/c_0)
t'$ for time, $\eta=a \eta\,'$ for surface elevation and $\varphi=(g l_x
a/c_0)\varphi\,'$ for velocity potential, where $c_0=\sqrt{g h_0}$ is the mean
wave speed and $g$ is the constant gravity. The quantity $\sigma = \sigma\,' /
(h_0\rho c_0^2)$ is the dimensionless Bond number, in which $\rho$ is the mass
density of the fluid and  $\sigma'$ is its surface tension, both of which are
taken to be constants. After dropping primes, this asymptotic expansion yields
the nondimensional Korteweg-de Vries (KdV) equation for the horizontal velocity
variable $u=\varphi_x(x,t)$ at {\it linear} order in the small
dimensionless ratios $\epsilon$ and $\delta^2$, as the left hand side of
\begin{equation}\label{kdv-eqn}
u_t+u_x+\frac{3\epsilon}{2}uu_x
+\frac{\delta^2}{6}(1-3\sigma)u_{xxx}
=O(\epsilon\delta^2)
\,.
\end{equation}
Here, partial derivatives are denoted using subscripts, and boundary
conditions are $u=0$ and $u_x=0$ at spatial infinity on the real line. The
famous  $sech^2(x-t)$ traveling wave solutions (the solitons) for KdV
(\ref{kdv-eqn}) arise in a balance between its (weakly) nonlinear steepening
and its third-order linear dispersion, when the quadratic terms in $\epsilon$
and $\delta^2$ on its right hand side are neglected.

In equation (\ref{kdv-eqn}), a normal form transformation due to Kodama
\cite{Ko1985} has been used to remove the other possible quadratic terms of
order $O(\epsilon^2)$ and $O(\delta^4)$. The remaining quadratic correction
terms in the KdV equation (\ref{kdv-eqn}) may be collected at order
$O(\epsilon\delta^2)$. These terms may be expressed, after introducing a
``momentum variable,'' 
\begin{equation}\label{mom-def}
m=u-\nu\delta^2u_{xx}
\,,
\end{equation}
and neglecting terms of {\it cubic} order in $\epsilon$ and $\delta^2$, as
\begin{equation}\label{b-eqn}
m_t+m_x+\frac{\epsilon}{2}(um_x+b\,mu_x)
+\frac{\delta^2}{6}(1-3\sigma)u_{xxx}
=0
\,.
\end{equation}
In the momentum variable $m=u-\nu\delta^2u_{xx}$, the parameter $\nu$ is given
by \cite{DuGoHo2001}
\begin{equation}\label{nu-def}
\nu=\frac{19-30\sigma-45\sigma^2}{60(1-3\sigma)}
\,.
\end{equation}
Thus, the effects of $\delta^2-$dispersion also enter the nonlinear terms.
After restoring dimensions in equation (\ref{b-eqn}) and rescaling velocity 
$u$ by $(b+1)$, the following ``$b-$equation" emerges,
\begin{equation}\label{dim-b-eqn}
m_t+c_0m_x
+um_x+b\,mu_x
+\Gamma u_{xxx}
=0
\,,
\end{equation}
where $m=u-\alpha^2u_{xx}$ is the dimensional momentum variable, and the
constants $\alpha^2$ and $\Gamma/c_0$ are squares of length scales.
When $\alpha^2\to0$, one recovers KdV from the $b-$equation (\ref{dim-b-eqn}),
up to a rescaling of velocity. Any value of the parameter $b\ne -1$ may be
achieved in equation (\ref{dim-b-eqn}) by an appropriate Kodama transformation
\cite{DuGoHo2001}. 

As we have emphasized, the values of the coefficients in the asymptotic
analysis of shallow water waves at quadratic order in their two small
parameters only hold, modulo the Kodama normal-form transformations. Hence,
these transformations may be used to advance the analysis and thereby gain
insight, by optimizing the choices of these coefficients. The freedom
introduced by the Kodama transformations among asymptotically equivalent
equations at quadratic order in $\epsilon$ and $\delta^2$ also helps to answer
the perennial question, ``Why are integrable equations so ubiquitous when one
uses asymptotics in modeling?''

\paragraph{Integrable cases of the $b-$equation (\ref{dim-b-eqn}).}
The cases $b=2$ and $b=3$ are special values, for which the $b-$equation
becomes a completely integrable Hamiltonian system. For $b=2$, equation
(\ref{dim-b-eqn}) specializes to the integrable CH equation of Camassa and Holm
\cite{CaHo1993}. The case $b=3$ in (\ref{dim-b-eqn}) recovers the integrable 
DP equation of Degasperis and Procesi \cite{DePr1999}. These two cases exhaust
the integrable candidates for (\ref{dim-b-eqn}), as was shown using Painlev\'e
analysis. The $b-$family of equations (\ref{dim-b-eqn})
was also shown in \cite{MiNo2002} to admit the symmetry conditions necessary
for integrability, only in the cases $b = 2$ for CH and $b = 3$ for DP. 

The $b-$equation (\ref{dim-b-eqn}) with $b=2$ was first derived in 
Camassa and Holm \cite{CaHo1993} by using asymptotic expansions directly in
the Hamiltonian for Euler's equations governing inviscid incompressible flow
in the shallow water regime. In this analysis, the CH equation was shown to
be bi-Hamiltonian and thereby was found to be completely integrable by the
inverse scattering transform (IST) on the real line. Reviews of IST may be
found, for example, in Ablowitz et al. \cite{AbCl1991}, Dubrovin {\it et al.}
\cite{DuNoKr1985}, Novikov {\it et al.}
\cite{NoMaPiZa1984}. For discussions of other related bi-Hamiltonian
equations, see \cite{DePr1999}.

Camassa and Holm \cite{CaHo1993} also discovered the remarkable peaked
soliton (peakon) solutions of (\ref{singlepeakon-soln},\ref{peakontrain-soln})
for the CH equation on the real line, given by (\ref{dim-b-eqn}) in the case
$b=2$. The peakons arise as solutions of (\ref{dim-b-eqn}), when $c_0 = 0$ and
$\Gamma = 0$ in the absence of linear dispersion.  Peakons move at a speed
equal to their maximum height, at which they have a sharp peak (jump in
derivative). Unlike the KdV soliton, the peakon speed is independent of its
width ($\alpha$).  Periodic peakon solutions of CH were treated in Alber {\it
et al.} \cite{AlCaFeHoMa1999+2001}. There, the sharp peaks of
periodic peakons were associated with billiards reflecting at  the boundary of
an elliptical domain. These billiard solutions for the periodic peakons arise
from  geodesic motion on a tri-axial ellipsoid, in the limit that one of its
axes shrinks to zero length. 

Before Camassa and Holm derived their shallow water equation in
\cite{CaHo1993}, a class of integrable equations existed, which was later
found to contain equation (\ref{dim-b-eqn}) with $b=2$. This class of
integrable equations was derived using hereditary symmetries in Fokas and
Fuchssteiner \cite{FoFu1981}. However, equation (\ref{dim-b-eqn}) was not
written explicitly, nor was it derived physically as a shallow water equation
and its solution properties for $b=2$ were not studied before Camassa and Holm
\cite{CaHo1993}. See Fuchssteiner \cite{Fu1996} for an insightful history of
how the shallow water equation (\ref{dim-b-eqn}) in the integrable case with
$b=2$ relates to the mathematical theory of hereditary symmetries. 

Equation (\ref{dim-b-eqn}) with $b=2$ was recently re-derived as a shallow
water equation by using asymptotic methods in three different approaches in
Dullin {\it et al.} \cite{DuGoHo2001}, in Fokas and Liu \cite{FoLi1996} and
also in Johnson \cite{Jo2002}. These three derivations all used different
variants of the method of asymptotic expansions for shallow water waves in the
absence of surface tension. Only the derivation in Dullin {\it et al.} 
\cite{DuGoHo2001} used the Kodama normal-form transformations to take
advantage of the non-uniqueness of the asymptotic expansion results at
quadratic order.

The effects of the parameter $b$ on the solutions of equation (\ref{dim-b-eqn})
were investigated in Holm and Staley \cite{HoSt2003a}, where $b$
was treated as a bifurcation parameter, in the limiting case when the linear
dispersion coefficients are set to $c_0 = 0$ and $\Gamma = 0$. This
limiting case allows several special solutions, including the peakons, in
which the two nonlinear terms in equation (\ref{dim-b-eqn}) balance each
other in the {\it absence} of linear dispersion.

\section{Peakons: Singular solutions without linear dispersion in one spatial
dimension}  Peakons were first found as singular soliton solutions of the
completely integrable CH equation. This is equation (\ref{dim-b-eqn}) with
$b=2$, now rewritten in terms of the velocity, as
\begin{eqnarray}\label{CH-u-eqn}
u_t+c_0u_x+3uu_x&+&\Gamma u_{xxx}
\nonumber\\
&=&\alpha^2(u_{xxt}+2u_xu_{xx}+uu_{xxx})
\,.
\end{eqnarray}
Peakons were found in \cite{CaHo1993} to arise in the absence of linear
dispersion. That is, they arise when $c_0=0$ and $\Gamma=0$ in CH
(\ref{CH-u-eqn}). Specifically, peakons are the individual terms in the peaked
$N-$soliton solution of CH (\ref{CH-u-eqn}) for its velocity,
\begin{equation}\label{CHpeakon-u-soln}
u(x,t)=\sum_{b=1}^N \, p_b(t)e^{-|x-q_b(t)|/\alpha}
\,,
\end{equation}
in the absence of linear dispersion.
Each term in the sum is a solition with a sharp peak at its maximum. Hence,
the name ``peakon.'' Expressed using its momentum,
$m=(1-\alpha^2\partial_x^2)u$, the peakon velocity solution
(\ref{CHpeakon-u-soln}) of dispersionless CH becomes a sum over a delta
functions, supported on a set of points moving on the real line. Namely, 
the peakon velocity solution (\ref{CHpeakon-u-soln}) implies
\begin{equation}\label{CHpeakon-m-soln}
m(x,t)=2\alpha\sum_{b=1}^N \, p_b(t)\delta(x-q_b(t))
\,,\end{equation}
because of the relation
$(1-\alpha^2\partial_x^2)e^{-|x|/\alpha}=2\alpha\delta(x)$. These
solutions satisfy the $b-$equation (\ref{dim-b-eqn}) for any value of $b$,
provided $c_0 = 0$ and $\Gamma = 0$. 

Thus, peakons are {\it singular momentum solutions} of the dispersionless
$b-$equation, although they are not stable for every value of $b$. From
numerical simulations \cite{HoSt2003a}, peakons are conjecture to
be stable for $b>1$. In the integrable cases $b = 2$ for CH and $b = 3$ for DP,
peakons are stable singular {\it soliton} solutions. The spatial velocity
profile $e^{-|x|/\alpha}/(2\alpha)$ of each separate peakon in
(\ref{CHpeakon-u-soln}) is the Green's function for the Helmholtz operator on
the real line, with vanishing boundary conditions at spatial infinity. Unlike
the KdV soliton, whose speed and width are related, the width of the peakon
profile is set by its Green's function, independently of its speed.

\paragraph{Integrable peakon dynamics of CH.} 
Substituting the peakon solution ansatz (\ref{CHpeakon-u-soln})
and (\ref{CHpeakon-m-soln}) into the dispersionless CH equation,
\begin{equation}\label{DCHmom-eqn}
m_t
+
um_x
+
2mu_x
=0
\,,\quad\hbox{with}\quad
m=u-\alpha^2u_{xx}
\,,
\end{equation}
yields {\it Hamilton's canonical equations} for the dynamics of the discrete
set of peakon parameters $p_a(t)$ and $q_a(t)$, 
\begin{equation}\label{Ham-peakon-eqn}
\dot{q}_a(t) = \frac{\partial h_N}{\partial p_a}
\quad\hbox{and}\quad
\dot{p}_a(t) = -\,\frac{\partial h_N}{\partial q_a}
\,,
\end{equation}
for $a=1,2,\dots,N$, with Hamiltonian given by \cite{CaHo1993},
\begin{equation}\label{H-peakon-ansatz}
h_N=\tfrac{1}{2}\sum_{a,b=1}^N p_a\,p_b\,e^{-|q_a-q_b|/\alpha}
\,.
\end{equation}
Thus, one finds that the points $x=q^a(t)$ in the peakon solution
(\ref{CHpeakon-u-soln}) move with the flow of the fluid velocity $u$ at
those points, since $u(q^a(t),t)=\dot{q}^a(t)$. This means the $q^a(t)$ are
{\it Lagrangian} coordinates. Moreover, the singular momentum solution ansatz
(\ref{CHpeakon-m-soln}) is the Lagrange-to-Euler map for an invariant
manifold of the dispersionless CH equation (\ref{DCHmom-eqn}). On this
finite-dimensional invariant manifold for the partial differential
equation (\ref{DCHmom-eqn}), the dynamics is canonically Hamiltonian. 

With Hamiltonian (\ref{H-peakon-ansatz}), the canonical
equations (\ref{Ham-peakon-eqn}) for the $2N$ canonically conjugate peakon
parameters $p_a(t)$ and $q_a(t)$ were interpreted in \cite{CaHo1993}
as describing {\it geodesic motion} on the
$N-$dimensional Riemannian manifold whose co-metric is
$g^{ij}(\{q\})=e^{-|q_i-q_j|/\alpha}$. Moreover, the canonical geodesic
equations arising from Hamiltonian (\ref{H-peakon-ansatz}) comprise an
integrable system for any number of peakons $N$. This integrable system was
studied in \cite{CaHo1993} for solutions on the real line, and in
\cite{AlCaFeHoMa1999+2001,McCo1999} and references therein, for spatially
periodic solutions.  \rem{The integrable solutions on the real line were
related to the Toda chain with open ends, in \cite{Va2003}.}

Being a completely integrable Hamiltonian soliton equation, the continuum CH
equation (\ref{CH-u-eqn}) has an associated isospectral eigenvalue problem,
discovered in \cite{CaHo1993} for any values of its dispersion parameters
$c_0$ and $\Gamma$. Remarkably, when $c_0 = 0$ and $\Gamma = 0$, this
isospectral eigenvalue problem has a purely {\it discrete} spectrum. Moreover,
in this case, each discrete eigenvalue corresponds precisely to the
time-asymptotic velocity of a peakon. This discreteness of the CH isospectrum
in the absence of linear dispersion implies that {\it only} the singular
peakon solutions (\ref{CHpeakon-m-soln}) emerge asymptotically in time, in the
solution of the initial value problem for the dispersionless CH equation 
(\ref{DCHmom-eqn}). This is borne out in numerical simulations of the
dispersionless CH equation (\ref{DCHmom-eqn}), starting from a smooth initial
distribution of velocity \cite{FrHo2001,HoSt2003a}. 

\begin{figure}[ht]
\begin{center}
\includegraphics[scale=0.75,angle=0]{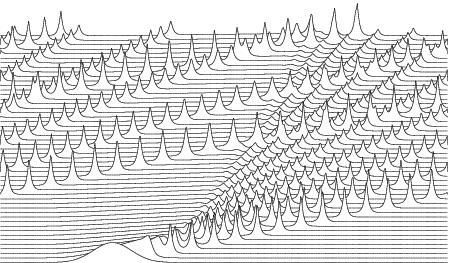}
\end{center}
\caption{
A smooth localized (Gaussian) initial condition for the CH equation breaks
up into an ordered train of peakons as time evolves (the time direction
being vertical). The peakon train eventually wraps around the periodic
domain, thereby allowing the leading peakons to overtake the slower
emergent peakons from behind in collisions that cause phase shifts, as
discussed in \cite{CaHo1993}.}
\label{peakon_figure}
\end{figure}

Figure \ref{peakon_figure}
shows the emergence of peakons from an initially Gaussian
velocity distribution and their subsequent elastic collisions in
a periodic one-dimensional domain.\footnote{The figures in this article  
were kindly supplied by Martin Staley} This
figure demonstrates that singular solutions dominate the initial
value problem and, thus, that it is imperative to go beyond
smooth solutions for the CH equation; the situation is similar
for the EPDiff equation.

\paragraph{Peakons as mechanical systems.} 
Being governed by canonical Hamiltonian equations, each $N-$peakon solution
can be associated with a mechanical system of moving particles. Calogero et
al. \cite{Ca1995} further extended the class of mechanical systems of this
type. The r-matrix approach was applied to the Lax pair formulation of the
$N-$peakon system for CH by Ragnisco and Bruschi 
\cite{RaBr1996}, who also pointed out the connection of this system with the
classical Toda lattice. A discrete version of the Adler-Kostant-Symes
factorization method was used by Suris \cite{Su1996} to study a
discretization of the peakon lattice, realized as a discrete integrable system
on a certain Poisson submanifold of $gl(N)$ equipped with an r-matrix Poisson
bracket. Beals {\it et al.} \cite{BeSaSm1999+2000} used the
Stieltjes theorem on continued fractions and the classical moment problem for
studying multi-peakon solutions of the CH equation. Generalized peakon systems
are described for any simple Lie algebra by Alber {\it et al.}
\cite{AlCaFeHoMa1999+2001}.

\paragraph{Pulsons: Generalizing the peakon solutions of the
dispersionless $b-$equation for other Green's functions.}  The 
Hamiltonian $h_N$ in equation (\ref{H-peakon-ansatz}) depends on the Green's
function for the relation between velocity $u$ and momentum $m$. However, the
singular momentum solution ansatz (\ref{CHpeakon-m-soln}) is {\it independent}
of this Green's function. Thus, as discovered in Fringer and Holm
\cite{FrHo2001}, 

\noindent
{\it The singular momentum solution ansatz
(\ref{CHpeakon-m-soln}) for the dispersionless equation,
\begin{equation}\label{FHmom-eqn}
m_t
+
um_x
+
2mu_x
=0
\,,\quad\hbox{with}\quad
u=g*m
\,,
\end{equation}
provides an invariant manifold on which canonical Hamiltonian dynamics occurs,
for {\it any choice} of the Green's function $g$ relating velocity $u$ and
momentum $m$ by the convolution $u=g*m$. }

The fluid velocity solutions corresponding to the singular momentum ansatz  
(\ref{CHpeakon-m-soln}) for equation (\ref{FHmom-eqn}) are the {\it pulsons}.
Pulsons are given by the sum over $N$ velocity profiles determined by the
Green's function $g$, as
\begin{equation}\label{pulson-u-soln}
u(x,t)=\sum_{a=1}^N \, p_a(t)g\big(x,q_a(t)\big)
\,.
\end{equation}
Again for (\ref{FHmom-eqn}), the singular momentum ansatz
(\ref{CHpeakon-m-soln}) results in a finite-dimensional invariant manifold of
solutions, whose dynamics is canonically Hamiltonian.  The Hamiltonian for the
canonical dynamics of the $2N$ parameters $p_a(t)$ and
$q_a(t)$ in the ``pulson'' solutions (\ref{pulson-u-soln}) of equation
(\ref{FHmom-eqn}) is 
\begin{equation}\label{H-pulson-ansatz}
h_N=\tfrac{1}{2}\sum_{a,b=1}^N p_a\,p_b\,g(q_a,q_b)
\,.
\end{equation}
Again for the pulsons, the canonical equations for the invariant manifold of
singular momentum solutions provide a phase-space description of geodesic
motion, this time with respect to the co-metric given by the Green's function
$g$. Mathematical analysis and numerical results for the dynamics of these
pulson solutions are given in \cite{FrHo2001}. These results describe how the
collisions of  pulsons (\ref{pulson-u-soln}) depend upon their shape.%

\paragraph{Compactons in the $1/\alpha^2\to0$ limit of CH.} As mentioned
earlier, in the limit that $\alpha^2\to0$, the CH equation (\ref{CH-u-eqn})
becomes the KdV equation. In the opposite limit that $1/\alpha^2\to0$ CH
becomes the Hunter-Zheng equation \cite{HuZh1994}
\[
\big(u_t+uu_x\big)_{xx} = \frac{1}{2}(u_x^2)_x
\hspace{2cm}(\hbox{Hunter-Zheng})
\]
This equation has ``compacton'' solutions, whose collision dynamics was
studied numerically and put into the present context in \cite{FrHo2001}. The
corresponding Green's function satisfies $-\partial_x^2g(x)=2\delta(x)$, so it
has the triangular shape, $g(x)=1-|x|$ for $|x|<1$, and vanishes otherwise,
for $|x|\ge1$. That is, the Green's function in this case has compact support;
hence, the name ``compactons'' for these pulson solutions, which as a
limit of the integrable CH equations are true solitons, solvable by IST. 

\paragraph{Pulson solutions of the dispersionless $b-$equation.}
Holm and Staley \cite{HoSt2003a} give the pulson solutions of the
traveling wave problem and their elastic collision properties for
the dispersionless $b-$equation,
\begin{equation}\label{disp-b-eqn}
m_t
+
um_x
+
b\,mu_x
=0
\,,
\quad\hbox{with}\quad
u=g*m
\,,
\end{equation}
with any (symmetric) Green's function $g$ and for any value of the parameter
$b$. Numerically, pulsons and peakons are both found to be stable for $b>1$,
\cite{HoSt2003a}. The reduction to {\it noncanonical} Hamiltonian
dynamics for the invariant manifold of singular momentum solutions
(\ref{CHpeakon-m-soln}) of the other integrable case $b=3$ with peakon Green's
function $g(x,y)=e^{-|x-y|/\alpha}$ is found in
\cite{DePr1999}.


\section{Euler-Poincar\'e theory in more dimensions}

\paragraph{Generalizing the peakon solutions of the CH equation to
higher dimensions.}  
In \cite{HoSt2003a}, weakly nonlinear analysis and the assumption of
columnar motion in the variational principle for Euler's equations are found
to produce the two-dimensional generalization of the dispersionless CH
equation (\ref{DCHmom-eqn}). This generalization is the Euler-Poincar\'e (EP)
equation
\cite{HoMaRa1998a} for the Lagrangian consisting of the kinetic energy,
\begin{equation}
\ell
=
\frac{1}{2}\int 
\Big[|{\mathbf{u}}|^2
+
\alpha^2\big(
{\rm div\,}{\mathbf{u}}
\big)^2
\Big]dxdy
\,,
\label{KE-def}
\end{equation} 
in which the fluid velocity ${\mathbf{u}}$ is a two-dimensional vector.
Evolution generated by kinetic energy in Hamilton's principle results in
geodesic motion, with respect to the velocity norm $\|{\mathbf{u}}\|$, which
is  provided by the kinetic energy Lagrangian. For ideal incompressible
fluids governed by Euler's equations, the importance of geodesic flow was
recognized by Arnold \cite{Ar1966} for the $L^2$ norm of the fluid velocity.
The EP equation generated by any choice of kinetic energy norm without
imposing incompressibility is called ``EPDiff,'' for ``Euler-Poincar\'e
equation for geodesic motion on the diffeomorphisms.''  EPDiff is given by
\cite{HoMaRa1998a}
\begin{equation}\label{EPDiff-eqn}
\Big(\frac{\partial}{\partial t}
+
{\mathbf{u}}\cdot\nabla\Big)
\mathbf{m}
+
\nabla \mathbf{u}^T\cdot
\mathbf{m}
+
\mathbf{m}
({\rm div\,}{\mathbf{u}})
=
0
\,,\end{equation}
with momentum density 
$\mathbf{m} 
= 
\delta\ell/\delta{\mathbf{u}}
\,,$
where $\ell=\frac{1}{2}\|{\mathbf{u}}\|^2$ is given by the kinetic energy,
which defines a norm in the fluid velocity $\|{\mathbf{u}}\|$, yet to be
determined. By design, this equation has no contribution from either potential
energy, or pressure. It conserves the velocity norm $\|{\mathbf{u}}\|$ given
by the kinetic energy. Its evolution describes  geodesic motion on the
diffeomorphisms with respect to this norm
\cite{HoMaRa1998a}. An alternative way of writing the EPDiff equation
(\ref{EPDiff-eqn}) in either two, or three dimensions is,
\begin{equation}\label{H1-EPcurl-eqn}
\frac{\partial}{\partial t}
\mathbf{m}
-
{\mathbf{u}}\times{\rm curl\,}{\mathbf{m}}
+
\nabla({\mathbf{u}}\cdot{\mathbf{m}})
+
\mathbf{m}
({\rm div\,}{\mathbf{u}})
=
0
\,.\end{equation}
This form of EPDiff involves all three differential operators, curl,
gradient and divergence. 
For the kinetic energy Lagrangian $\ell$ given in (\ref{KE-def}), which is a
norm for {\it irrotational} flow (with ${\rm curl\,}{\mathbf{u}}=0$), we have 
the EPDiff equation (\ref{EPDiff-eqn}) with momentum $\mathbf{m}  = 
\delta\ell/\delta{\mathbf{u}}
=
\mathbf{u}
-
\alpha^2\nabla({\rm div\,}{\mathbf{u}})$. 

EPDiff (\ref{EPDiff-eqn}) may also be written intrinsically as
\begin{equation}\label{EPDiffeo-eqn}
\frac{\partial}{\partial t}
\frac{\delta\ell}{\delta {\mathbf{u}}}
=
-
{\,\rm ad}^*_{\mathbf{u}}
\frac{\delta\ell}{\delta {\mathbf{u}}}
\,,
\end{equation}
where ad$^*$ is the $L^2$ dual of the ad-operation (commutator) for vector
fields. See \cite{ArKh1998,MaRa1999} for additional discussions of the
beautiful geometry underlying this equation.

\paragraph{Reduction to the dispersionless CH equation in 1D.}
In one dimension, the EPDiff equation (\ref{EPDiff-eqn}-\ref{EPDiffeo-eqn})
with Lagrangian
$\ell$ given in (\ref{KE-def}) simplifies to the dispersionless CH equation
(\ref{DCHmom-eqn}). The dispersionless limit of the CH equation
appears, because we have ignored potential energy and pressure.

\paragraph{Strengthening the kinetic energy norm to allow for circulation.}
The kinetic energy Lagrangian (\ref{KE-def}) is a norm for irrotational flow,
with ${\rm curl\,}{\mathbf{u}}=0$. However, inclusion of rotational
flow requires the kinetic energy norm to be strengthened to the $H_\alpha^1$
norm of the velocity, defined as
\begin{eqnarray}
\ell
&=&
\frac{1}{2}
\int 
\bigg[|{\mathbf{u}}|^2
+
\alpha^2\big(
{\rm div\,}{\mathbf{u}}
\big)^2
+
\alpha^2\big(
{\rm curl\,}{\mathbf{u}}
\big)^2
\bigg]dxdy
\nonumber\\
&=&
\frac{1}{2}
\int 
\bigg[|{\mathbf{u}}|^2
+
\alpha^2|\nabla{\mathbf{u}}|^2
\bigg]dxdy
=
\frac{1}{2}
\|{\mathbf{u}}\|^2_{H_\alpha^1}
\,.
\label{H1-eqn}
\end{eqnarray} 
Here we assume boundary conditions that give no contributions upon  
integrating by parts. The corresponding EPDiff equation is (\ref{EPDiff-eqn})
with $\mathbf{m}
\equiv
\delta \ell/\delta \mathbf{u}
=
\mathbf{u}
-
\alpha^2\Delta{\mathbf{u}}
\,.$
This expression involves inversion of the familiar {\it Helmholtz} operator in
the (nonlocal) relation between fluid velocity and momentum density. The
$H_\alpha^1$ norm $\|{\mathbf{u}}\|^2_{H_\alpha^1}$ for the kinetic energy
(\ref{H1-eqn}) also arises in three dimensions for turbulence modeling based
on Lagrangian averaging and using Taylor's hypothesis that the turbulent
fluctuations are ``frozen'' into the Lagrangian mean flow
\cite{FoHoTi2001}. 

\paragraph{Generalizing the CH peakon solutions to $n$ dimensions.}
Building on the peakon solutions (\ref{CHpeakon-u-soln}) for the CH equation
and the pulsons (\ref{pulson-u-soln}) for its generalization to other
traveling-wave shapes in \cite{FrHo2001}, Holm and Staley
\cite{HoSt2003a} introduced the following measure-valued singular
momentum solution ansatz for the
$n-$dimensional solutions of the EPDiff
equation (\ref{EPDiff-eqn}):
\begin{equation}\label{m-ansatz}
\mathbf{m}(\mathbf{x},t)
=
\sum_{a=1}^N\int\mathbf{P}^a(s,t)\,
\delta\big(\,\mathbf{x}-\mathbf{Q}^a(s,t)\,\big)ds.
\end{equation}
These singular momentum solutions, called ``diffeons,'' are vector density
functions supported in ${\mathbb{R}}^n$ on a set of $N$ surfaces (or curves)
of  codimension $(n-k)$ for $s\in {\mathbb{R}}^{k}$ with $k<n$.  They may, for
example, be supported on sets of points (vector peakons, $k=0$),
one-dimensional filaments (strings, $k=1$), or two-dimensional surfaces
(sheets, $k=2$) in three dimensions. 

\begin{figure}[ht]
\centering
\includegraphics[scale=0.75,angle=0]{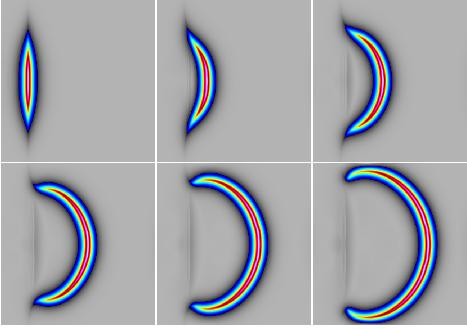}
\caption{
A peakon segment of finite length is initially moving
rightward (East). Because its speed vanishes at its ends and it has fully 
two-dimensional spatial dependence, it expands into a peakon ``bubble''
as it propagates.  (The colors indicate speed: red is highest, yellow is
less, blue low, grey zero.) }
\label{2Dstrip_plate}
\end{figure}

Figure \ref{2Dstrip_plate}
shows the results for the EPDiff equation when a straight peakon segment of
finite length is created initially moving rightward (East). Because of  
propagation along the segment in adjusting to the condition of zero
speed at its ends and finite speed in its interior, the initially straight
segment expands outward as it propagates and curves into a peakon ``bubble.''

\begin{figure}[ht]
\begin{center}
\includegraphics[scale=0.75,angle=0]{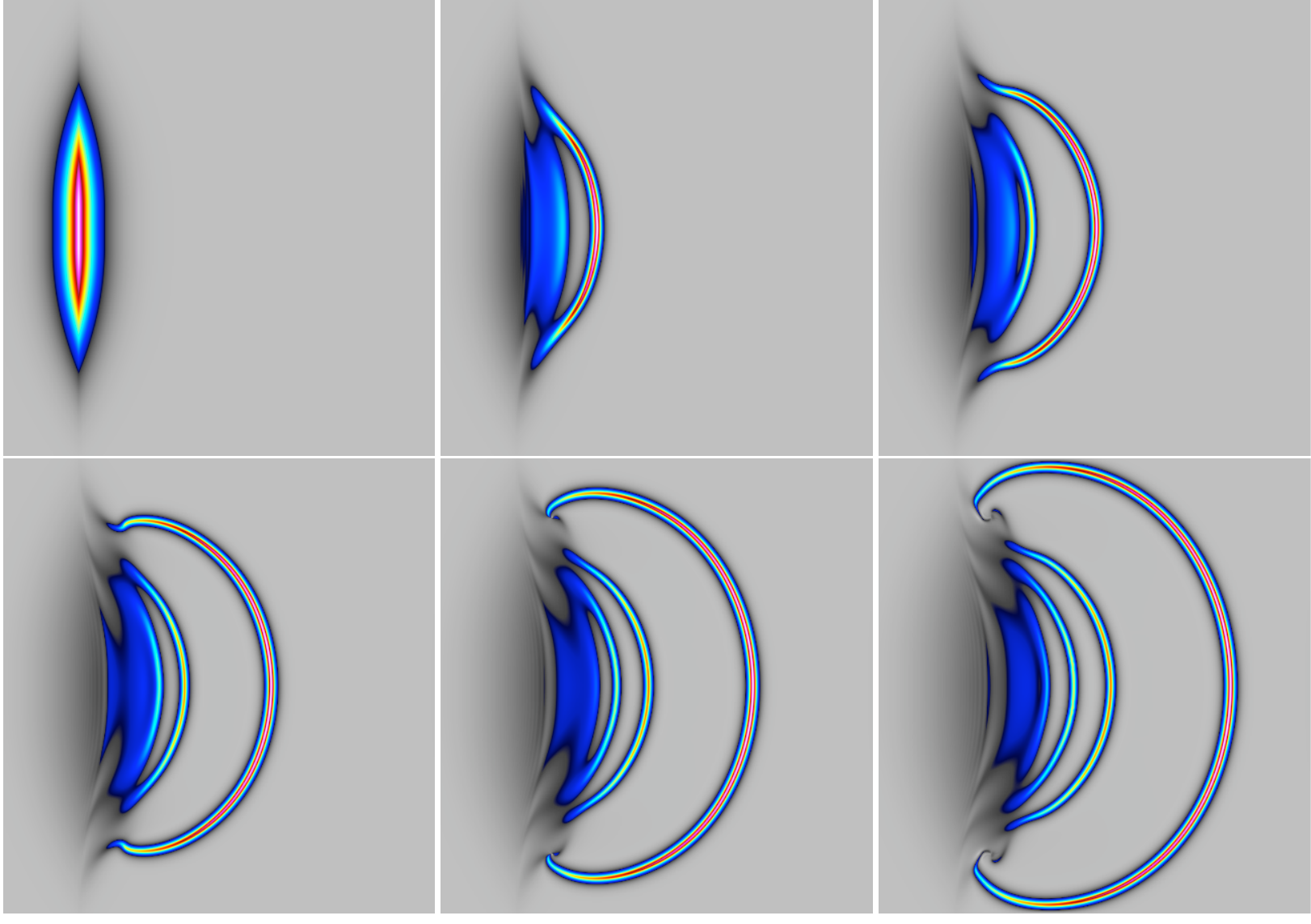}
\end{center}
\caption{
An initially straight segment of velocity distribution whose exponential
profile is wider than the width $\alpha$ for the peakon solution will break
up into a train of curved peakon ``bubbles,'' each of width $\alpha$. This
example illustrates the emergent property of the peakon solutions in two
dimensions. }
\label{2Dstrip_plate-multi}
\end{figure}

Figure \ref{2Dstrip_plate-multi} shows an
initially straight segment whose velocity distribution is exponential in
the transverse direction, but is wider than $\alpha$ for the peakon solution.
This initial velocity distribution evolves under EPDiff to separate into a
train of curved peakon ``bubbles,'' each of width $\alpha$. This
example illustrates the emergent property of the peakon solutions in two
dimensions. This phenomenon is observed in nature, for example, as trains of
internal wave fronts in the south China Sea \cite{Liu-etal[1998]}.

\begin{figure}[ht]
\begin{center}
\includegraphics[scale=0.75,angle=0]{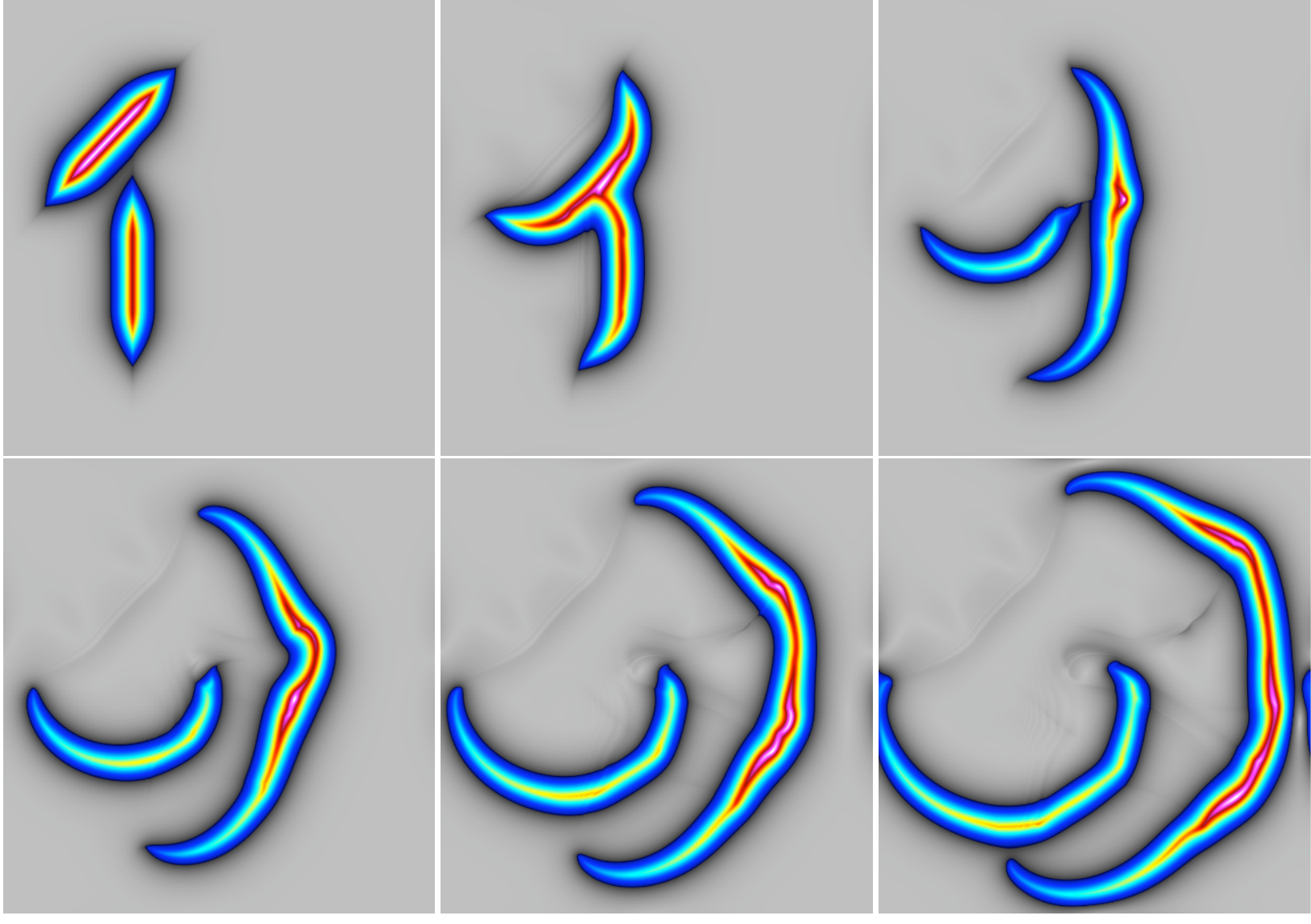}
\end{center}
\caption{
A single collision is shown involving reconnection as the faster
peakon segment initially moving Southeast along the diagonal expands, curves 
and obliquely overtakes the slower peakon segment initially moving rightward
(East). This reconnection illustrates one of the collision rules for the
strongly two-dimensional EPDiff flow. }
\label{2Dstrip_skew}
\end{figure}

 Substitution of the singular momentum solution ansatz
(\ref{m-ansatz}) into the EPDiff
equation (\ref{EPDiff-eqn}) implies the following 
integro-partial-differential equations (IPDEs) for the evolution of
the parameters $\{{\mathbf{P}}\}$ and $\{{\mathbf{Q}}\}$,
\begin{eqnarray}
\frac{\partial }{\partial t}\mathbf{Q}^a (s,t)
&=&
\sum_{b=1}^{N} \int\mathbf{P}^b(s^{\prime},t)\,
G(\mathbf{Q}^a(s,t)-\mathbf{Q}^b(s^{\prime},t)\,\big)ds^{\prime}
\,,\label{IntDiffEqn-Q}
\nonumber\\
\frac{\partial }{\partial t}\mathbf{P}^a (s,t)
&=&
-\,\!\!\sum_{b=1}^{N} \int
\big(\mathbf{P}^a(s,t)\!\cdot\!\mathbf{P}^b(s^{\prime},t)\big)
\\&&\hspace{5mm}
\frac{\partial }{\partial \mathbf{Q}^a(s,t)}
G\big(\mathbf{Q}^a(s,t)-\mathbf{Q}^b(s^{\prime},t)\big)\,ds^{\prime}
\,.
\nonumber
\end{eqnarray}
%
Importantly for the interpretation of these solutions, the coordinates
$s\in {\mathbb{R}}^{k}$ turn out to be {\it Lagrangian} coordinates. The 
velocity field corresponding to the momentum solution ansatz
(\ref{m-ansatz}) is given by
\begin{eqnarray}\label{u-ansatz}
\mathbf{u}(\mathbf{x},t)
&=&
G*\mathbf{m}
\nonumber\\
&=&
\sum_{b=1}^N\int\mathbf{P}^b(s^{\prime},t)\,
G\big(\,\mathbf{x}-\mathbf{Q}^b(s^{\prime},t)\,\big)ds^{\prime}
\,,
\end{eqnarray}
for $\mathbf{u}\in{\mathbb{R}}^n$.
When evaluated along the curve $\mathbf{x}=\mathbf{Q}^a(s,t)$, this
velocity satisfies,
\begin{eqnarray}\label{Qdot-ansatz}
\mathbf{u}(\mathbf{Q}^a(s,t),t)
&=&
\sum_{b=1}^N\int\mathbf{P}^b(s^{\prime},t)\,
G\big(\,\mathbf{Q}^a(s,t)
-\mathbf{Q}^b(s^{\prime},t)\,\big)ds^{\prime}
\nonumber \\
&=&
\frac{\partial\mathbf{Q}^a(s,t)}{\partial t}
\,.
\end{eqnarray}
Consequently, the lower-dimensional support sets defined on
$\mathbf{x}=\mathbf{Q}^a(s,t)$ and parameterized by coordinates
$s\in{\mathbb{R}}^{k}$ move with the fluid velocity. This means the
$s\in{\mathbb{R}}^{k}$ are Lagrangian coordinates. Moreover, equations
(\ref{IntDiffEqn-Q}) for the evolution of these support sets are canonical
Hamiltonian equations,
\begin{equation}
\label{IntDiffEqns-Ham}
\frac{\partial }{\partial t}\mathbf{{Q}}^a (s,t)
=
\frac{\delta H_N}{\delta \mathbf{P}^a}
\,,\qquad
\frac{\partial }{\partial t}\mathbf{{P}}^a (s,t)
=
-\,\frac{\delta H_N}{\delta \mathbf{Q}^a}
\,.
\end{equation}
The corresponding Hamiltonian function 
$H_N:({\mathbb{R}}^n\times {\mathbb{R}}^n)^{N}\to {\mathbb{R}}$ is,
\begin{eqnarray} 
H_N &=& \frac{1}{2}\!\int\!\!\!\!\int\!\!\sum_{a\,,\,b=1}^{N}
\big(\mathbf{P}^a(s,t)\cdot\mathbf{P}^b(s^{\prime},t)\big)
\nonumber\\&&\hspace{1cm}
G\big(\mathbf{Q}^a(s,t),\mathbf{Q}_{\,j}(s^{\prime},t)\big)
\,ds\,ds^{\prime}
\,.\label{H_N-def}
\end{eqnarray}
This is the Hamiltonian for geodesic motion on the cotangent bundle of
a set of curves $\mathbf{Q}^a(s,t)$ with respect to the metric given by
$G$. This dynamics was investigated numerically in 
\cite{HoSt2003a} to which we refer for more details of
the solution properties. One important result found {\it numerically} in 
\cite{HoSt2003a} is that only codimension-one singular momentum solutions 
appear to be stable under the evolution of the EPDiff equation. Thus, we have

{\it Stability for codimension-one: the singular momentum solutions of
EPDiff are stable, as points on the line (peakons), as curves in the plane
(filaments, or wave fronts), or as surfaces in space (sheets). }

Proving this stability result analytically remains an outstanding problem. The
stability of peakons on the real line is proven in \cite{CoSt2000}.

\begin{figure}[ht]
\begin{center}
\includegraphics[scale=0.75,angle=0]{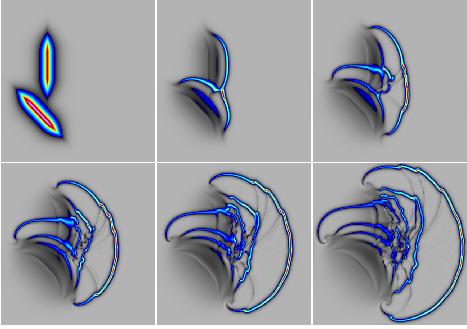}
\end{center}
\caption{
A series of multiple collisions is shown involving reconnections as the
faster wider peakon segment initially moving Northeast along the diagonal
expands, breaks up into a wave train of peakons, each of which propagates,
curves and obliquely overtakes the slower wide peakon segment initially moving
rightward (East), which is also breaking up into a train of wavefronts. In
this series of oblique collision, the now-curved peakon filaments exchange
momentum and reconnect several times. }
\label{2Dstrip_skew-multi}
\end{figure}

\paragraph{Reconnections in oblique overtaking collisions of peakon wave fronts.}  
Figures \ref{2Dstrip_skew} and \ref{2Dstrip_skew-multi}  show 
results of oblique wave front collisions producing reconnections for the
EPDiff equation in two dimensions. Figure \ref{2Dstrip_skew} shows a single
oblique overtaking collision, as a faster expanding peakon wave front
overtakes a slower one and reconnects with it at the collision point. Figure
\ref{2Dstrip_skew-multi} shows a series of reconnections involving the
oblique overtaking collisions of two trains of curved peakon filaments, or
wave fronts.

\paragraph{The peakon reduction is a momentum map.} As shown in
\cite{HoMa2004}, the singular solution ansatz (\ref{m-ansatz}) is a momentum 
map from the cotangent bundle of the smooth embeddings of lower dimensional
sets ${\mathbb{R}}^s\in{\mathbb{R}}^n\,,$ to the dual of the Lie algebra of
vector fields defined on these sets. (Momentum maps for Hamiltonian
dynamics are reviewed in \cite{MaRa1999}, for example.)  This geometric
feature underlies the remarkable reduction properties of the EPDiff equation,
and it also explains why the reduced  equations must be Hamiltonian on the
invariant manifolds of the singular solutions; namely, because momentum maps
are Poisson maps. This geometric feature also underlies the singular momentum
solution (\ref{m-ansatz}) and its associated velocity (\ref{u-ansatz})
which generalize the peakon solutions, both to higher dimensions and to
arbitrary kinetic energy metrics. The result that the singular solution ansatz
\textup{(\ref{m-ansatz})} is a momentum map helps to organize the theory, to
explain previous results and to suggest new avenues of exploration.

\paragraph{Acknowledgments.} I am grateful to R. Camassa, J. E. Marsden, T. S.
Ratiu and A. Weinstein for their collaboration, help and inspiring discussions
over the years. I also thank M. F. Staley for providing the figures
obtained from his numerical simulations in our collaborations. US DOE provided
partial support, under contract W-7405-ENG-36 for Los Alamos National
Laboratory, and Office of Science ASCAR/AMS/MICS.


\end{document}